%% file: arxiv_ezz.tex
\documentclass[%
preprint,
 amsmath,amssymb,
 aps,
 pra,
]{revtex4-1}

\usepackage{graphicx}
\usepackage{color,xcolor,soul,pstricks}
\usepackage{dcolumn}
\usepackage{bm}

\newcommand{\mkdf}[2]{\frac{\partial{#1}}{\partial{#2}}}
\newcommand{\mkddf}[2]{\frac{\partial^2{#1}}{\partial{#2}^2 }}

\newcommand{\mkDf}[2]{\frac{d{#1}}{d{#2}}}

\begin{document}


\title{Generalized spheroidal wave equation for real \\ and complex valued parameters. An algorithm based on the analytic derivatives for the eigenvalues }

\author{Mykhaylo V. Khoma}
\affiliation{%
Max Planck Institute for the Physics of Complex Systems, N\"othnitzer Str. 38, 
Dresden-01187, Germany  \\
Institute of Electron Physics,
National Academy of Sciences of Ukraine, Uzhhorod, Ukraine 
}%





\date{\today}

\begin{abstract}

This paper presents a new approach for the computation of eigenvalues of the generalized spheroidal wave equations. The novelty of the present method is in the use of the analytical derivatives of the eigenvalues to minimize losses in accuracy. The derivatives are constructed in the form of three-term recurrent relations within the method of continued fractions associated with the corresponding spheroidal wave equation. Very accurate results for the eigenvalues are obtained for a wide range of the parameters of the problem. As an illustrative example, the electronic energies and the separation constants are computed for various electronic states and geometries of selected ($\rm{H}_2^{+}$, $\rm{HeH}^{2+}$, and $\rm{BH}^{5+}$) quasimolecular systems. The computations for high lying ${}^{2}\Sigma$ electronic states of $\rm{H}_2^{+}$ up to very large internuclear separations ($ \leq 1.7 \times 10^5$ au) are presented.
Also presented the computations for the eigenvalues of the generalized spheroidal wave equations with complex valued parameters. The agreement between the obtained results and the results of other authors is discussed.


\end{abstract}

\pacs{Valid PACS appear here}
\maketitle


\section{Introduction}\label{sec:intro}


Generalized spheroidal wave equations (GSWEs) play a fundamental role in many research 
areas such as atomic and molecular physics \cite{BrownCarr-book, Kereselidze-epjd2024a, Kereselidze-epjd2024b, LKhJ-jpb, Bondar07, Ponce-NIMB97, A-Frecon-jphysb-89, Hadinger-jphysb-89, EuSink83, Rost-Hyperfine}, signal processing and electrodynamics \cite{Hrycak-IEEE, Osipov-book, Deng08, LeWei-ElMag-book}, gravitation and cosmology 
\cite{SlavLay-book, Kereselidze-epjd2025, Teukolsky-2015, Fiziev-2010, Leaver86}, etc. Several approaches to determine the solutions for the GSWEs  have been proposed in the literature \cite{Peek65,Peek71, Hunter-67, Hodge, Power73, Leaver86, RankinThorson, Ponomarev-76, KPS76-book, Macek-sturmian-PRA, AbramovKurochkin-2006, Figueiredo-16, Figueiredo-2002, Bonorino-48, Bonorino-49, Scott-06, Liu92, Falloon-03, Falloon-thesis}. 

However, despite many efforts, methods of solution for these equations certainly need further development. 
This seems to be related to the fact that the standard algorithms for their evaluation 
encounter numerical difficulties in practical computations for very large  or complex valued parameters of the GSWE, as was noted in \cite{Liu92,Falloon-03,Falloon-thesis,YanPengGong-PRE09,Charles-98,Li-PRE-1998}. Finding efficient algorithms to compute GSWEs is still an area of ongoing research and a number of new studies have appeared in recent years
\cite{CPC-2023-Fursa, Mitnik21, Seri-15, CPC-spheroidal-2014, Cook-14, Schmid-2023, Schmid-2025}.

The approach to solving the GSWEs boils down to finding their eigenvalues via non-linear search algorithms starting from (sufficiently good) initial guesses. The difficulty mostly resides in finding such good trial eigenvalues for their subsequent non-linear refinement. In the present study we eliminate this difficulty by using analytical derivatives of the sought-for eigenvalues with respect to the distance $R$ between the foci of the spheroidal system of coordinates and the boundary conditions at $R=0$. 

Perhaps the most notable example of the application of GSWEs is 
the two-Coulomb-center problem $eZ_1Z_2$, i.e., the problem of solving the Schr\"odinger equation for the motion of an electron in the field of two Coulomb centers with charges $Z_1$ and $Z_2$ at a fixed distance $R$ from each other. In fact, mathematically the two-Coulomb-center problem is equivalent to two coupled GSWEs \cite{Leaver86,Liu92,SlavLay-book}. This paper is concerned with the accurate numerical solutions of GSWEs and applications of such solutions to the $eZ_1Z_2$ problem. 

The structure of the paper is the following. We provide the necessary mathematical preliminaries and formulate the problem in Sec. \ref{sec:formulation}. In Sec. \ref{sec:master_equation} we present the main 
mathematical result of this work, viz., analytic representations for the derivatives of the eigenvalues of the $eZ_1Z_2$ problem with respect to the internuclear distance. These analytical derivatives are used in Sec. \ref{sec:numeric} in numerical calculations for the $eZ_1Z_2$ problem, including the symmetric 
$\rm{H}_2^+$ and asymmetric ($\rm{HeH}^{2+}$ and $\rm{BH}^{5+}$) cases. In the same section we present our results for the highly excited ${}^{2}\Sigma$ electronic states of $\rm{H}_2^+$ up to very large internuclear distances ($\sim 10^5$ au). 
Continuum electronic states for $eZ_1Z_2$ problem are considered in subsection \ref{subsec:continuum}. Calculations of the eigenvalues for complex valued parameters are presented in subsection \ref{subsec:complex}.
We conclude in Sec. \ref{sec:conclusion}. In the appendix we provide auxiliary information needed for section \ref{sec:formulation}.

Unless otherwise stated, atomic units ($\hbar=m_e=e=1$) are used throughout this paper.

\section{Formulation of the problem}\label{sec:formulation}

\subsection{The GSWE and the two-Coulomb-center problem}

The generalized spheroidal wave equation can be given in the following form \cite{Leaver86}:
\begin{equation}\label{def-GSWE}
t(t-t_0) \mkddf {y}{t} + \left( B_1 + B_2 t \right) \mkdf {y}{t} 
 + [ \omega^2 t (t-t_0) - 2 \alpha \omega(t-t_0) + B_3  ] y = 0,
\end{equation}
where $B_1$, $B_2$, $B_3$, $\alpha$, $\omega$, and $t_0$ are constants. The intervals of the independent variable $t$ interesting for physical applications are both $[ 0 \leq t \leq t_0 ]$ and $[t_0 \leq t < \infty)$. 

In view of application to the two-Coulomb-center problem, we use the prolate spheroidal coordinates $(\xi,\eta,\varphi)$. 
If $Z_1$ and $Z_2$ are the charges placed at a fixed distance $R$ along a line defining the $z$ axis and 
$r_1$ and $r_2$ are the distances of an electron from charges $Z_1$ and $Z_2$, respectively, then $\xi$, $\eta$, and $\varphi$ are defined as follows:
\[
\xi = (r_1 + r_2)/R \, , \quad \eta=(r_1-r_2)/R  \, ,  \quad \varphi=\arctan(y/x),
\]
\begin{equation}\label{def-spheroidal-coords}
1 \leq \xi < \infty \, ,   \quad   -1 \leq \eta \leq 1 \, ,   \quad   0 \leq \varphi < 2\pi,
\end{equation}
where $\varphi$ is the azimuthal angle with respect to the $z$ axis. The Schr\"odinger  equation for the $e Z_1 Z_2$ problem reads
\begin{equation}\label{def-ShEq}
\nabla^2  \Psi + 2\big( E -V \big) \Psi(\vec{r};R) = 0,
\end{equation}
where $\Psi(\vec{r};R)$ is the electronic wave function, $\vec{r} = (x,y,z)$ is the position vector of the electron, $E$ is the electronic energy, and $V =-Z_1/r_1 - Z_2/r_2$ is the electron-nuclei potential. In the coordinates \eqref{def-spheroidal-coords} the potential $V$ reads
\begin{equation}\label{def-V-potential}
V = -\frac{2 \xi Z_{+} }{ R (\xi^2 - \eta^2)} - 
\frac{ 2 \eta Z_{-} }{R (\xi^2 - \eta^2)},
\end{equation}
where $Z_{+}=Z_1+Z_2$ and $Z_{-}=Z_2-Z_1$. For the sake of  definiteness, we assume $Z_2 \geq Z_1$. 
 
Owing to the form of the potential \eqref{def-V-potential}, equation \eqref{def-ShEq} is separable in the prolate spheroidal coordinates \eqref{def-spheroidal-coords}. Therefore, the wave function $\Psi$ can be factorized as follows:  
\begin{equation}\label{spheroidal-func}
\Psi(\vec{r};R) =  \Pi(\xi) \, \Xi (\eta)  \frac{ e^{ \, \pm \, im\varphi}}{ \sqrt{ 2 \pi}}.
\end{equation}
The functions $\Pi(\xi)$ and $ \Xi (\eta)$ satisfy the following system of differential equations: 
\begin{equation}\label{def0-radial-eq}
\Big[  L_{\xi} 
+ \Big( \lambda^{(\xi)}  + c^2 (\xi^{2}-1)
+ a \, \xi- \frac{ m^{2} } { \xi ^{2}-1 } \Big) \Big] \Pi(\xi)=0,
\end{equation}
\begin{equation}\label{def0-angular-eq}
\Big[ L_{\eta}
+ \Big( -\lambda^{(\eta)} + c^2 (1-\eta ^{2}) +b \, \eta -\frac{m^{2}} {1-\eta^{2}} \Big) \Big] \Xi (\eta)=0,
\end{equation}
where 
\begin{equation}
L_{\xi} =  \partial_{\xi}  (\xi ^{2}- 1) \partial_{\xi}, \quad 
L_{\eta} = \partial_{\eta} (1-\eta^{2}) \partial_{\eta},
\end{equation}
\begin{equation}\label{def-ab}
a=R Z_+, \quad  b=R Z_-,
\end{equation}
$\lambda^{(\xi)}$, $\lambda^{(\eta)}$, and $m$ are the separation constants ($m$ is assumed to be a non-negative integer and is referred to as the modulus of the magnetic quantum number \cite{Power73}).  In the case $E<0$ it is convenient to introduce the following parameters: $\varepsilon = \sqrt{-2E}$ and $p= \varepsilon R /2$.

Equations \eqref{def0-radial-eq} and \eqref{def0-angular-eq} are GSWEs. This can be demonstrated by using the substitutions $\Pi(\xi) = (\xi^2 - 1)^{m/2} f(\xi)$ and $ \Xi (\eta) = (1- \eta^2)^{m/2} g(\eta)$ for \eqref{def0-radial-eq} and \eqref{def0-angular-eq}, respectively. The differential equations for $f(\xi)$ and $g(\eta)$ will take the form of equation \eqref{def-GSWE} if we let $\xi=t-1 $ for the function $f(\xi)$ and $\eta=t-1$ for the function $g(\eta)$, respectively \cite{Leaver86, Liu92}. The parameters in \eqref{def-GSWE} then read as follows: $t_0=2$, $\omega = c$, $\alpha= - R(Z_1 \pm Z_2)/2c$, $B_1 = -2(m+1)$, $B_2 = 2(m+1)$, and $B_3 = \lambda + m(m+1) + R(Z_1 \pm Z_2)$. 

Let us note here that the GSWE is also referred to as the confluent Heun equation (for details see  \cite{Bonorino-48,Bonorino-49,SlavLay-book}). Also note that equation \eqref{def0-radial-eq} is referred to as the Coulomb Spheroidal Wave Equation CSWE in \cite{Schmid-2023}, whereas the term GSWE refers to a slightly more general equation (see also \cite{KPS76-book}).

The form of Eqs. \eqref{def0-radial-eq} and \eqref{def0-angular-eq} is the most suitable for treating
the continuum states. Here we first consider the discrete spectrum ($E<0$) of the $eZ_1Z_2$ problem, which is described by Eqs. \eqref{def0-radial-eq} and \eqref{def0-angular-eq} subject to the formal substitution $c^2 \to - p^2$ (i.e. $c^2 = ER^2/2$). 
The differential equations for $\Pi(\xi)$ and $\Xi (\eta)$ then reads
\begin{equation}\label{def-radial-eq}
\Big[  L_{\xi} + \Big( \lambda^{(\xi)} - p^{2}(\xi^{2}-1)
+ a \, \xi- \frac{ m^{2} } { \xi ^{2}-1 } \Big) \Big] \Pi(\xi)=0, 
\end{equation}
\begin{equation}\label{def-angular-eq}
\Big[ L_{\eta}
 + \Big( -\lambda^{(\eta)} -p^{2}(1-\eta ^{2}) +b \, \eta -\frac{m^{2}} {1-\eta^{2}} \Big) \Big] 
\Xi (\eta)=0. 
\end{equation}
The functions $\Pi(\xi)$ and $\Xi (\eta)$ satisfy the following boundary conditions:
\begin{equation}\label{boundary-cond1}
|\Pi(1)| < \infty, \quad \lim_{\xi \to \infty} \Pi(\xi) = 0,
\end{equation}
\begin{equation}\label{boundary-cond2}
| \Xi (\pm 1)| < \infty.
\end{equation}
Under the conditions \eqref{boundary-cond1} and \eqref{boundary-cond2} equations 
\eqref{def-radial-eq} and \eqref{def-angular-eq} 
each define a Sturm-Liouville problem with the spectral parameters
$\lambda^{(\xi)}$ and $\lambda^{(\eta)}$, respectively \cite{Janev-book}. The eigenfunctions and eigenvalues of the problem are characterized by the quantum numbers $k$ and $m$ for 
$\Pi_{km}(\xi)$ and $\lambda_{km}^{(\xi)}$, and $q$ and $m$ for $\Xi_{qm}(\eta)$ and $\lambda_{qm}^{(\eta)}$. The numbers $k$ and $q$ are equal to the numbers of nodes of the eigenfunctions $\Pi_{km}(\xi)$ and $\Xi_{qm}(\eta)$, respectively \cite{Hunter-67,Power73,Hodge}.
The quantum numbers $(k,q,m)$ are related to the "united atom" quantum numbers $(N, l, m)$ as follows: $N=k+q+m+1$, $l=q+m$ ($0 \leq l \leq N-1$, $0 \leq m \leq l$). 
The system of coupled equations \eqref{def0-radial-eq} and \eqref{def0-angular-eq} represents the solution of the Schr\"odinger equation \eqref{def-ShEq} under the condition that the separation constants 
$\lambda_{km}^{(\xi)}$ and $\lambda_{qm}^{(\eta)}$ are equal: 
\begin{equation}\label{lambda-xi-eta}
\lambda^{(\xi)}_{km}(p,a) = \lambda^{(\eta)}_{qm}(p,b) \equiv \lambda. 
\end{equation} 
Equation \eqref{lambda-xi-eta} posses a unique solution for $\lambda$ and $p$ at fixed values of $a$ and $b$ (or, equivalently, $R$, $Z_1$, and $Z_2$) and quantum numbers $k$, $l$(=$q+m$), and $m$, i.e.,
$\lambda \equiv \lambda_{lm}(R)$ and $p \equiv p_{kqm}(R)$.
Usually, the $\lambda_{lm}(R)$ and $p_{kqm}(R)$ are computed in an iterative manner starting from the known analytical expressions in the united atom limit:
\begin{equation}\label{UA-ini}
\lambda_{lm}(R \to 0) = -l(l+1), \quad 
p_{kqm}(R \to 0) = 0.
\end{equation} 
Using the relation between $p$ and $E$ [see after equation \eqref{def-ab}] one obtains the discrete energy spectrum $E_{kqm}(R)$. 

\subsection{The algorithm to compute the eigenvalues }

The solutions to the equations \eqref{def0-radial-eq} and \eqref{def0-angular-eq} for the discrete spectrum are expanded in terms of some basis functions $u_s(\xi)$ and $v_s(\eta)$ as follows:
$\Pi(\xi) = \sum_{s=0}^{\infty} c_s^{(\xi)} u_s(\xi)$ and $\Xi (\eta)=   \sum_{s=0}^{\infty} c_s^{(\eta)} v_s (\eta)$.
%
%
Substitution of these expansions into the equations for $\Pi(\xi)$ and $\Xi (\eta)$  lead to the following three-term recurrent relations for coefficients $c_s^{(\xi)}$ and $c_s^{(\eta)}$
\begin{equation}\label{rec-rel-common}
\alpha_s^{(\xi,\eta)} \, c_{s+1}^{(\xi,\eta)} - \beta_s^{(\xi,\eta)} \, c_s^{(\xi,\eta)} +
\gamma_s^{(\xi,\eta)} \, c_{s-1}^{(\xi,\eta)} = 0,  
\end{equation}
with boundary conditions $c^{(\xi, \eta)}_{-1}=0$.
The traditional choice for the radial basis $u_s(\xi)$ is due to Jaffe \cite{Jaffe}: 
\begin{equation}\label{Jaffe-expans} 
\Pi(\xi) = (\xi^2-1)^{m/2} (\xi+1)^{\sigma} e^{-p \, \xi} 
\sum_{s=0}^{\infty} c^{(\xi)}_s \left( \frac{\xi-1}{\xi+1} \right)^s,
\end{equation}
where $\sigma=({a}/{2p})-m-1$. 

The angular basis $v_s(\eta)$ depends on the range of the internuclear distance $R$. For not very large distances ($R \lesssim$ 10 a.u.) the expansion of
$ \Xi (\eta)$ in the associated Legendre functions $P_{s+m}^{\,m}(\eta)$ is used \cite{KPS76-book,Janev-book}:
\begin{equation}\label{ANG1-expans}
\Xi_{qm}(\eta) = \left( 1- \eta^2 \right)^{m/2}e^{-p \eta} \sum_{s=0}^{\infty} 
c^{(\eta)}_s P_{s+m}^{\,m}(\eta).
\end{equation}
For large $R$ the expansion of $\Xi(\eta)$ in terms of $P_{s+m}^{\,m}(\eta)$
is found to converge badly \cite{Pyzunina-68}. In this case, one uses 
the following basis functions:
\begin{equation}\label{ANG2}
v^{\pm}_{s}(\eta) = \left( 1- \eta^2 \right)^{m/2} \exp(\mp p  \eta) (1 \pm \eta)^{s}.
\end{equation}
For the computation of the eigenvalues 
both the $v_s^{+}$ and $v_s^{-}$ are equivalent. For the computation of the wave functions, however, the basis $v_s^{-}$ is preferred for the $\eta$-interval $[0,1]$ and $v_s^{+}$ for the $\eta$-interval $[-1,0]$, vice versa.   
The explicit form of coefficients $\alpha_s^{(\xi, \eta)}$, $\beta_s^{(\xi, \eta)}$, and 
$\gamma_s^{(\xi, \eta)}$ can be find elsewhere (see, e.g., \cite{Janev-book, Power73, Leaver86, Hodge, Hunter-67} and references therein). For the sake of completeness we provide these coefficients in the appendix.

In the framework of the method of continued fractions (CFs) the eigenvalues $\lambda^{(\xi)}$ and $\lambda^{(\eta)}$ are calculated at fixed $a$ and $b$ by finding the roots of the following system of coupled characteristic equations \cite{Janev-book,Leaver86,Liu92,Power73}:
\begin{equation}
\label{chain-frac-xi}
\frac{F^{(\xi)}}{Q^{(\xi)}} \equiv  \beta^{(\xi)}_{0}-
\frac{\alpha^{(\xi)}_{0}\gamma^{(\xi)}_{1}}{\beta^{(\xi)}_{1}-}\,
\frac{\alpha^{(\xi)}_{1}\gamma^{(\xi)}_{2}}{\beta^{(\xi)}_{2}-}\,
\frac{\alpha^{(\xi)}_{2}\gamma^{(\xi)}_{3}}{\beta^{(\xi)}_{3}- \ldots} = 0,
\end{equation}
\begin{equation}
\label{chain-frac-eta}
\frac{F^{(\eta)}}{Q^{(\eta)}} \equiv  \beta^{(\eta)}_{0}-
\frac{\alpha^{(\eta)}_{0}\gamma^{(\eta)}_{1}}{\beta^{(\eta)}_{1}-}\,
\frac{\alpha^{(\eta)}_{1}\gamma^{(\eta)}_{2}}{\beta^{(\eta)}_{2}-}\,
\frac{\alpha^{(\eta)}_{2}\gamma^{(\eta)}_{3}}{\beta^{(\eta)}_{3}- \ldots} = 0.
\end{equation}
Here $F^{(\xi,\eta)}$ and $Q^{(\xi,\eta)}$ are the numerators and denominators of the corresponding CF. Obviously,  \eqref{chain-frac-xi} and \eqref{chain-frac-eta}  are equivalent to the following equations: 
\begin{equation}\label{Pxi-zero-eq}
F_{N_\xi}^{(\xi)} (p, a, \lambda^{(\xi)} ) = 0,
\end{equation}
\begin{equation}\label{Peta-zero-eq}
F_{N_\eta} ^{(\eta)} (p, b, \lambda^{(\eta)} ) = 0.
\end{equation}
In practice, CF is truncated at the (sufficiently large) $N$-th term. Here the subscripts $N_\xi$ and $N_\eta$ denote the number of terms (the length) of the corresponding CF.  
Expressions for $F_{N_\xi}^{(\xi)}$ and $F_{N_\eta}^{(\eta)}$ can be obtained by means of the recurrence relations \cite{Power73,Gautschi,Wall-book}
\begin{equation}\label{Pxi-Peta-recurr}
F_j^{(\xi,\eta)} = \beta_j^{(\xi,\eta)} \, F_{j-1}^{(\xi,\eta)} - 
\alpha_{j-1}^{(\xi,\eta)} \gamma_{j}^{(\xi,\eta)} \, F_{j-2}^{(\xi,\eta)},
\end{equation}
with boundary conditions $F^{(\xi,\eta)}_{-2} = 0$ and $F^{(\xi,\eta)}_{-1} = 1$.  For brevity, further we omit the subscripts $N_{\xi}$ and $N_{\eta}$ in notations for $F_{N_\xi}^{(\xi)}$ and 
$F_{N_\eta}^{(\eta)}$. Where necessary these values will be specified explicitly.

The separation constants $\lambda^{(\xi)}$ and $\lambda^{(\eta)}$ and the parameter $p$ must be so chosen to satisfy simultaneously equations \eqref{Pxi-zero-eq} and \eqref{Peta-zero-eq} and 
condition \eqref{lambda-xi-eta} at fixed value of $R$ and quantum numbers ${k,q,m}$. The electronic energy $E$ is found directly from the parameter $p$.


Solving system of Eqs. \eqref{Pxi-zero-eq} and \eqref{Peta-zero-eq} 
involves a nonlinear root finding using, e.g., the Newton-Raphson  
algorithm, which starts from some trial values ${p}_t$ and ${\lambda_t}$.
The bottleneck of this approach is that the method is rather demanding of the quality of these trial parameters. The straightforward application of the root search routines, even for the one-dimensional case related to the continuum spectra of the problem, becomes very impractical for large parameters $c$, $l$, and $m$ \cite{Liu92}. The situation is even more complex for two-dimensional case of the system of equations \eqref{Pxi-zero-eq} and \eqref{Peta-zero-eq}.

Another method, due to Hunter and Pritchard \citep{Hunter-67} and Hodge \citep{Hodge}, consists of direct diagonalization of the matrix equation obtained from the three-term recurrence relations in Eq. \eqref{Pxi-Peta-recurr}. However, as was noted in Refs. \citep{Liu92,Falloon-03}, 
this method is also inefficient for obtaining highly accurate eigenvalues.
The method of Killingbeck \cite{A-Frecon-jphysb-89,Hadinger-jphysb-89} also relies strongly on the availability of sufficiently accurate trial eigenvalues (see discussion in \citep{Liu92}).

In this work we propose a new way of obtaining high-precision eigenvalues 
of the $eZ_1 Z_2$ problem (and related GSWEs)  
through accurate numerical integration of the system of coupled differential equations for $\lambda(R)$ and $p(R)$ [or $\lambda(R)$ and $E(R)$] with the boundary conditions \eqref{UA-ini}.
The needed analytical expressions for the derivatives $\mkDf{\lambda(R)}{R}$, $\mkDf{E(R)}{R}$, and $\mkDf{p(R)}{R}$, will be constructed in the next section. We note that $\lambda$, $p$, and  $E$ are smooth functions of $R$, and hence we can safely use a high-order numerical algorithm for the present task. The resulting eigenvalues are then used as starting guesses for the Newton-Raphson refinement.

\section{Derivation of the equations for $\bf dE/dR$, $\bf dp/dR$, and $\bf d\lambda/dR$}\label{sec:master_equation}

In this section we derive the expressions for $\mkDf{E}{R}$, $\mkDf{p}{R}$, and 
$\mkDf{\lambda}{R}$ employing the chain rule and implicit differentiation methods. 
For the sake of brevity we have omitted the indices $k$, $q$, and $m$ in notations for $E$, $p$, and $\lambda$ where this is not confusing. We adopt the following abbreviations for the partial derivatives with respect to the arbitrary
variable $x$, i.e.:
\begin{equation}\label{notation-drv}
F^{(\xi,\eta)}_x \equiv  \mkdf{F^{(\xi,\eta)}}{x}, \quad   
\lambda^{(\xi,\eta)}_x \equiv  \mkdf{\lambda^{(\xi,\eta)}}{x}.
\end{equation}

Let us start from the derivation of the expression for $\mkDf{E}{R}$. The full differentials of the functions $F^{(\xi)}(p,a,\lambda)$ and $F^{(\eta)}(p,b,\lambda)$  with respect to the independent variable $R$ reads
\begin{equation}\label{dP-xi-dR}
dF^{(\xi)} = \left( \frac{\varepsilon}{2} F_{p}^{(\xi)} + Z_{+} F^{(\xi)}_{a} +   
F^{(\xi)}_{\lambda} \lambda^{(\xi)}_{R} \right) dR,
\end{equation}
\begin{equation}\label{dP-eta-dR}
dF^{(\eta)} = \left( \frac{ \varepsilon }{2} F_p^{(\eta)} + Z_{-} F_b^{(\eta)} +   
F_{\lambda}^{(\eta)}  \lambda_{R}^{(\eta)} \right) dR.
\end{equation}
In computation of the $dF^{(\xi)}$ and $dF^{(\eta)}$ we have used the explicit dependence of the variables $a$, $b$, and $p$ on $R$ (see section \ref{sec:formulation} A).

By equating $dF^{(\xi)}$ and $dF^{(\eta)}$ to zero [see Eqs. \eqref{Pxi-zero-eq} and \eqref{Peta-zero-eq}] we obtain directly from \eqref{dP-xi-dR} and \eqref{dP-eta-dR}
\begin{equation}\label{DLxiDR}
\lambda_{R}^{(\xi)} = \left(  F_{\lambda}^{(\xi)}  \right) ^{-1} 
\left( -\frac{\varepsilon }{2} F_{p}^{(\xi)} - Z_{+} F_{a}^{(\xi)} \right),
\end{equation}
\begin{equation}\label{DLetaDR}
\lambda_{R}^{(\eta)} = \left(  F_{\lambda}^{(\eta)}  \right) ^{-1} 
\left( -\frac{\varepsilon }{2} F_{p}^{(\eta)} - Z_{-} F_{b}^{(\eta)} \right).
\end{equation}
In the same manner, we can represent the full differentials $dF^{(\xi)}$ and $dF^{(\eta)}$ 
with respect to the independent variable $E$, namely
\begin{equation}\label{dP-xi-dE}
dF^{(\xi)} = \left( -\frac{R}{ 2 \varepsilon }  F_{p}^{(\xi)}  + 
F_{\lambda}^{(\xi)} \lambda_{E}^{(\xi)}  \right) dE,
\end{equation}
\begin{equation}\label{dP-eta-dE}
dF^{(\eta)} = \left( -\frac{R}{ 2 \varepsilon }  F_{p}^{(\eta)}  + 
F_{\lambda}^{(\eta)}  \lambda_{E}^{(\eta)}  \right) dE,
\end{equation}
which lead to the following relations:
\begin{equation}\label{DLxiDE}
\lambda_{E}^{(\xi)} = 
\frac{R}{ 2 \varepsilon } F_{p}^{(\xi)} \left( F_{\lambda}^{(\xi)} \right)^{-1},
\end{equation}
\begin{equation}\label{DLetaDE}
\lambda_{E}^{(\eta)} = 
\frac{R}{ 2 \varepsilon } F_{p}^{(\eta)} \left( F_{\lambda}^{(\eta)} \right)^{-1}.
\end{equation}
By taking the full differential of the expression \eqref{lambda-xi-eta} with respect to the independent variables $E$ and $R$, we obtain (as the derivative of an implicit function)
\begin{equation}\label{dEdR-general}
\frac{dE }{dR} = \left( \lambda^{(\xi)}_{R} -\lambda^{(\eta)}_{R} \right)
\left( \lambda^{(\eta)}_{E} -\lambda^{(\xi)}_{E} \right)^{-1}.
\end{equation}
Substituting now Eqs. \eqref{DLxiDR}, \eqref{DLetaDR}, \eqref{DLxiDE}, and \eqref{DLetaDE} 
into Eq. \eqref{dEdR-general}, we arrive at the final result 
%
\begin{equation}\label{dEdR-res}
\frac{dE (E,\lambda,R) }{dR}= \frac{\varepsilon^2}{R} + 
\frac{2\varepsilon 
\left( Z_{-}F_{b}^{(\eta)} F_{\lambda}^{(\xi)} - Z_{+} F_{a}^{(\xi)} F_{\lambda}^{(\eta)} \right)}
{R \left( F_{p}^{(\eta)} F_{\lambda}^{(\xi)} - F_{p}^{(\xi)} F_{\lambda}^{(\eta)}  \right)}.
\end{equation}
In numerical applications, however, it is more practical to use a pair of the derivatives 
$\mkDf{\lambda}{R}$ and $\mkDf{p}{R}$. The expression for $\mkDf{p}{R}$ can be obtained in the same way, and is given by
\begin{equation}\label{dpdR}
\frac{ dp (p,\lambda,R) } {dR}= \frac{ Z_{-} F_{b}^{(\eta)} F_{\lambda}^{(\xi)} -
 Z_{+}F_{a}^{(\xi)} F_{\lambda}^{(\eta)} }
{F_{p}^{(\xi)} F_{\lambda}^{(\eta)} - F_{p}^{(\eta)} F_{\lambda}^{(\xi)}  }.
\end{equation}

In a similar manner we can obtain the derivative $\mkDf{\lambda}{R}$. Again, let us write the full differentials $dF^{(\xi)}$ and $dF^{(\eta)}$ of the functions $F^{(\xi)}(p,a,\lambda)$ and $F^{(\eta)}(p,b,\lambda)$ in a general form: 
\begin{equation}\label{dPxi-typ2}
dF^{(\xi)} \equiv F_p^{(\xi)} dp +  F_{a}^{(\xi)} da +
F_{\lambda}^{(\xi)} d\lambda = 0,
\end{equation}
\begin{equation}\label{dPeta-typ2}
dF^{(\eta)} \equiv F_p^{(\eta)} dp +  F_{b}^{(\eta)} db +
F_{\lambda}^{(\eta)} d\lambda = 0.
\end{equation}
From Eq. \eqref{dPxi-typ2} it follows that
\begin{equation}
dp = - \bigg( F_a^{(\xi)} da + F_{\lambda}^{(\xi)} d\lambda \bigg)
\left( F_p^{(\xi)}  \right) ^{-1},
\end{equation}
which after substitution into Eq. \eqref{dPeta-typ2} leads to an auxiliary relation
\begin{equation}\label{DL-auxil}
\frac{
F_p^{(\eta)} 
}
{
F_p^{(\xi)} 
}
\big( F_a^{(\xi)} Z_{+} dR + 
F^{(\xi)}_{\lambda} d\lambda \big) =
F_b^{(\eta)} Z_{-} dR +  F^{(\eta)}_{\lambda} d\lambda.
\end{equation}
From the relation  \eqref{DL-auxil} we finally obtain
\begin{equation}\label{dLdR}
\frac{d \lambda (p,\lambda,R) }{dR} = 
\frac{
Z_{+} F_p^{(\eta)} F_a^{(\xi)} -
Z_{-} F_b^{(\eta)} F_p^{(\xi)}  
}
{
F_{\lambda}^{(\eta)} F_p^{(\xi)} -
{F_p^{(\eta)}} F_{\lambda}^{(\xi)} 
}.
\end{equation}

Thus, we obtained a system of the first-order differential equations \eqref{dpdR} and \eqref{dLdR} (or, alternatively, \eqref{dEdR-res} and \eqref{dLdR}). Solving this system is equivalent to finding the eigenvalues $p_{kqm}$ and $\lambda_{qm}$ of the GSWEs 
\eqref{def-radial-eq} and \eqref{def-angular-eq}.

\section{Details of numerical computations and examples}
\label{sec:numeric}

To illustrate the proposed algorithm, we have computed the electronic energies and separation constants
for the problem $eZ_1Z_2$ and represented here some examples. 
To compute $p(R)$ and $\lambda(R)$, we solve the system of algebraic equations \eqref{Pxi-zero-eq} and \eqref{Peta-zero-eq} by a Newton-Raphson method \cite{NumRecipes} with the initial guesses  
$\bar{p}$ and $\bar{\lambda}$ provided by solving the system of differential equations \eqref{dpdR} and \eqref{dLdR} at each propagation step of $R$. The numerical solution of the system of equations \eqref{dpdR} and \eqref{dLdR} obtained by using the 9th order explicit Runge-Kutta method \cite{RK-9}. 
The Newton-Raphson refinement require the knowledge of the Jacobian of the system (see Chapter 9.6 of \cite{NumRecipes}) 
\begin{equation}
J = \begin{tabular}{|cc|}
$ {F_p^{(\xi)}}$ & ${F_{\lambda}^{(\xi)}}$ \\ 
$ F_p^{(\eta)}$ & ${F_{\lambda}^{(\eta)}}$ \\ 
\end{tabular}~.
\end{equation}
The required expressions for $F_{p,\lambda}^{(\xi,\eta)}$ were computed by means of explicit analytic differentiation of the corresponding recurrence relations \eqref{Pxi-Peta-recurr}. 
In all our computations, we set $N_{\xi}$ and $N_{\eta}$ to be equal ($N_{\xi} = N_{\eta}$). 
Going further, we denote by $N_{\rm{ RK}}$ and $N_{\rm{NR}}$ the lengths of the continued fractions used for the Runge-Kutta propagation and for the Newton-Raphson refinement, respectively. For quantum numbers $(k,q) \leq 5$ and for the internuclear distances in the range $0.01 \leq R \leq 10$ a.u., it is enough to set $N_{\rm{RK}} = 80$ and $N_{\rm{NR}} = 160$ to ensure an accuracy of at least 28 digits for the $E_{kqm}$ and $\lambda_{kqm}$. 
In the present calculations, we have chosen to use 
$N_{\rm{RK}} =100$ and $N_{\rm{NR}}=300$, except stated explicitly. Also we set $\Delta R = 0.01$ a.u. for the Runge-Kutta step in $R$. Three, or at most four Newton steps, were sufficient to achieve the desired accuracy for all computations in this work. The present algorithm has been implemented in \textsc{fortran}-{\footnotesize 90} language. All computations of eigenvalues are carried out in quadruple precision arithmetic.

\subsection{Hydrogen molecular ion $\mathrm{H}_2^{+}$}

One of the important applications of GSWEs is the study of the hydrogen molecular ion $\mathrm{H}_2^+$, which plays a fundamental role in quantum chemistry, plasma physics, atomic scattering theory, astrophysics, and astronomical spectroscopy \cite{Nakashima-08,Nakashima-13,astro-2017,Turbiner-16,Greene18}, to cite a few. It further provides an excellent playground for testing a variety of numerical algorithms.

We have applied our method for computation of the electronic $E_{kqm}(R)$ and the Born-Oppenheimer (BO) potential energy curves (PECs) $E_{Nlm}^{\rm{BO}}(R)$ for the ground and some excited electronic states
of $\mathrm{H}_2^{+}$. We note that our results extend previous calculations in the literature.
Here and further on we denote the BO PECs with a united atom notation for quantum numbers. The distance $R_0=2.0$ a.u. is used as a benchmark internuclear separation. 
As a test of the present algorithm we also calculated the equilibrium distances $R_e$ (positions of the minimum of the BO PECs) for the electronic ground $1s\sigma_g$ and excited $2p\sigma_u$ and
$3d\sigma_g$ states. To compute the $R_e$ we gradually decrease the step in $R$, until the minimum was reached. For the $1s \sigma_g$  state we obtained $R_e =  1.997~193~319~969~9921$ a.u., which is very close to the value of $R_e$ reported in Ref. \cite{Nakashima-13}. However, the corresponding BO energy: $-$0.602~634~619~106~539~878~727~562~156~289~94 is noticeably different from that reported in  \cite{Nakashima-13} which, since our results on electronic energies are otherwise in excellent agreement with earlier computations from the same authors and of Ishikawa in \cite{Nakashima-08}, we suspect is due to a misprint.
We obtained the equilibrium distances of $R_e = 12.546~ 083~ 658~ 617~457$ a.u. with the BO energy:  
$-$0.500 060 790 563 912 563 640 093 300 090 05 a.u. for the $2p \sigma_u$ state, and $R_e$ = 8.834 164 503 179 2004 a.u. with the BO energy: $-$0.175 049 035 895 464 389 091 719 403 786 08 a.u. for the 
$3d \sigma_g$ state. Numerical results are summarized in Table \ref{table-01} (for the sake of compactness, we present the $E_{3d \sigma}^{\rm{BO}}$ energy with fewer digits than it was reported in \cite{Nakashima-13}).


\begin{table}[ht]
\caption{\label{table-01}%
The electronic energies $E_{kqm}$ and the BO energies $E_{Nlm}^{\rm{BO}}$ for the ground
$1s \sigma_{g}$ and some excited states of $\rm{H}_2^+$ computed at $R=2.0$ a.u.
and at specified equilibrium distances $R_e$. All values are given in atomic units (a.u.).
}
\begin{ruledtabular}
\renewcommand{\arraystretch}{0.7}
\begin{tabular}{lll}
& $R$ &  
\multicolumn{1}{c}{\textrm{energy}} \\
\colrule
&& \multicolumn{1}{c}{electronic state $1s\sigma_{g}$ } \\
$E_{000}$           &2.0& $-$1.102 634 214 494 946 461 508 968 945 318 34 \\
Ref. \cite{Nakashima-08} &2.0& $-$1.102 634 214 494 946 461 508 968 945 \\
%
{\textrm BO} & $R_e$\footnote{$R_e$ = 1.997 193 319 969 992 1}  
&$-0.602~634~ 619~ 106~ 539~ 878~ 727~ 562~ 156~289~94$  \\
Ref. \cite{Nakashima-13} & $R_e$\footnote{$R_e$ =  1.997 193 319 969 992 9} 
& $-0.602~634~619~106~539~869~378$ \\

&& \multicolumn{1}{c}{electronic state $2s \sigma_g$} \\
$E_{100}$&2.0& $-$0.360 864 875 339 503 845 038 699 751 181 {75} \\
Ref. \cite{ChinPhysB-20} &2.0& $-$0.360 864 875 339 503 845 038 699 8 \\

&& \multicolumn{1}{c}{electronic state $2p \sigma_u$} \\
$E_{010}$  & 2.0& $-$0.667 534 392 202 382 930 361 970 211 492 {12} \\  
Ref. \cite{Nakashima-08} &2.0& $-$0.667 534 392 202 382 930 361 970 211 49\\

\textrm{BO}  & $R_e$\footnote{$R_e$ = 12.546 083 658 617 457}
 & $-$0.500 060 790 563 912 563 640 093 300 090 05\\
Ref. \cite{Turbiner-16} &  $R_e$\footnote{$R_e$ = 12.545 25}  & $-$0.500 060 790 55 \\

&& \multicolumn{1}{c}{electronic state $2p \pi_u$} \\
$E_{001}$ &2.0& $-$0.428 771 819 895 856 436 313 960 091 139 85 \\
Ref. \cite{ChinPhysB-20} &2.0&$-$0.428 771 819 895 856 436 313 960 1\\

&& \multicolumn{1}{c}{electronic state $3p \sigma_u$} \\
$E_{110}$  &2.0& $-$0.255 413 165 086 484 561 417 250 236 137 06 \\
Ref. \cite{ChinPhysB-20} & 2.0 & $-$0.255 413 165 086 484 561 417 250 2 \\

&& \multicolumn{1}{c}{electronic state $3d \sigma_g$} \\
\textrm{BO} &  $R_e$\footnote{$R_e$ = 8.834 164 503 179 200 4}  & $-$0.175 049 035 895 464 389 091 719 403 786 08  \\ 
Ref. \cite{Nakashima-13}  & $R_e$\footnote{$R_e$ = 8.834 164 503 179 200 609 421 533 521}
 &$-$0.175 049 035 895 464 389 091 719 403 786 08 \\

&& \multicolumn{1}{c}{electronic state $4p \pi_u$} \\
BO & 2.0 & $+$0.384 084 709 963 404 777 081 630 847 953 27 \\
Ref. \cite{Tao-PRA09}  & 2.0 & $+$0.384 084 709 963 \\

\end{tabular}
\end{ruledtabular}
\end{table}

\subsubsection{The case of small $R$}

It is well known that Jaffe's expansion for the radial solution $\Pi(\xi)$ converges
slowly for very small $R$ (see, e.g., Ref. \cite{Nickel} and references therein). Therefore, calculation of the electronic energies of ${\mathrm{H}_2}^{+}$ at the small internuclear separations is a good test for the accuracy of the present method.

To verify our results we use an asymptotic (at $R \to 0$) representation for the energy of the $1s\sigma_g$ state. To the best of our knowledge, the most accurate asymptotic expression available for
$\rm{H}_2^+(1s \sigma_g)$ is the expansion to $O(R^{20})$ order obtained in Ref. \cite{Nickel}, whose result ensures precision of at least 34 decimal digits for the electronic energy at the internuclear distances $ \lesssim 0.005$ a.u.  
Our results and the result obtained from the asymptotic expansion of \cite{Nickel} are given 
in Table \ref{table-02}. The computational time increases rapidly with increasing 
the number of terms $N_{\rm{RK}}$ and $N_{\rm{NR}}$ of the continued fractions. To provide some information about the overall computational cost, the total elapsed time, including the time for one step of Runge-Kutta propagation and three Newton-Raphson refinements, is also reported in Table \ref{table-02}. Notice, however, that even for very moderate values of parameters $N_{\rm{NR}}$ and $N_{\rm{RK}}$, the result is accurate to 10 decimal digits.

\begin{table*}
\caption{\label{table-02} Convergence with increasing $N_{\rm{RK}}$ and $N_{\rm{NR}}$ of the electronic energy for the ground $1s \sigma_g$ state of $\rm{H}_2^+$ at $R=0.005$ a.u. The reference analytic asymptotic result from Ref. \cite{Nickel} is given at the bottom line of the table.
}
\begin{ruledtabular}
\begin{tabular}{ccll}
$N_{\rm{RK}}$ &  $N_{\rm{NR}}$  & energy (a.u.) & CPU time  \\
&&&  (min:sec)\footnote{Calculations were performed on a single core at 3.3 GHz.}\\ \hline
10 & 20    & $-$1.999 933 998 2 & 00:01\\
40 & 60    &$-$1.999 933 998 24 & 00:01\\
100  & 200 &$-$1.999 933 998 241 6   & 00:04\\
100  & 800    & $-$1.999 933 998 241 654 7 & 00:40 \\ 
100  & 2000   & $-$1.999 933 998 241 654 732 489 & 04:04 \\
100  & 4000   & $-$1.999 933 998 241 654 732 490 695 & 17:26\\
100 & 8000  &$-$1.999 933 998 241 654 732 490 695 394 316   & 64:28 \\
100 & 9000 &$-$1.999 933 998 241 654 732 490 695 394 316 08 & 81:25 \\
Ref. \cite{Nickel} && $-$1.999 933 998 241 654 732 490 695 394 316 079 & \\
\end{tabular}
\end{ruledtabular}
\end{table*}

\subsubsection{The case of large $R$}

Calculation of the BO energies at large internuclear distances is also a good test for the accuracy. We computed the $E_{1s\sigma}^{\rm{BO}}(R)$ energy of $\rm{H}_2^+$ at a sufficiently large $R$ and compared the results with the reference asymptotic values. 
The asymptotic expression for the degenerate at $R \to \infty$ energies of the $1s \sigma_g$ and $2p \sigma_u$ states of $\rm{H}_2^+$ reads \cite{Coulson}
\begin{equation}\label{Uasymp-R-large}
U(R) = -\frac{1}{2} -\frac{9}{4 R^4} - \sum_{n=6}^{10} \frac{C_n}{ R^{n}},
\end{equation}
where $C_6={15}/{2}$, $C_7 = { 213}/{4}$, $C_8 = {7755}/{64}$, $C_9 = {1773}/{2}$, and $C_{10} = {86049}/{16}$. As demonstrated in Table \ref{table-03}, our calculations are in excellent agreement with the
reference analytic results obtained from Eq. \eqref{Uasymp-R-large}. Due to the asymptotic nature of the expansion \eqref{Uasymp-R-large}, the values of $R \geq $ 2000 a.u. are required to ensure the precision of at least 32 digits for Eq. \eqref{Uasymp-R-large}. 
The exchange interaction, which decays exponentially at $R \to \infty$, can be totally neglected for such large values of $R$. Note the increasing values of $\Delta(R)$ (see Table \ref{table-03}) with decreasing of $R$ due to loss of precision of the asymptotic expansion \eqref{Uasymp-R-large}. Table \ref{table-04} gives the present results for the BO energies and $\lambda_{N l m}(R)$ for some excited states of $\rm{H}_2^+$ at large internuclear distances compared to the reference values from Ref. \cite{Peek71}. For $R<50$ a.u. we set $N_{\rm{RK}}=160$ and $N_{\rm{NR}}=600$, whereas for $R > 50 $ a.u. we set $N_{\rm{RK}}=240$ and $N_{\rm{NR}}=1000$.

\begin{table*}
\caption{\label{table-03} Convergence between the numerical $E_{1s\sigma}^{\rm{BO}}(R)$ and the asymptotic $U(R)$ energies for the ground $1s \sigma_g$ state of $\rm{H}_2^+$ with internuclear distance $R$. All values are given in atomic units. The numbers in parentheses denotes the powers of ten to be multiplied. Notation: $\Delta(R)= |E_{1s\sigma}^{\rm{BO}}(R) - U(R)| $. 
}
\begin{ruledtabular}
\begin{tabular}{clr}
$R$   & $E_{1s\sigma}^{\rm{BO}}(R)$   & $\Delta(R)$  \\
\hline
800      & $-$0.500 000 000 005 493 192 927 374 174 204 454 & 4.1$(-28)$\\
1000     & $-$0.500 000 000 002 250 007 553 372 063 701 698 & 5.1$(-29)$\\          
2000     & $-$0.500 000 000 000 140 625 117 603 990 689 269 & 6.5$(-32)$\\
3000     & $-$0.500 000 000 000 027 777 788 090 210 557 531 & 1.3$(-33)$ \\ 
\end{tabular}
\end{ruledtabular}
\end{table*}

\begin{table}[b]
\caption{\label{table-04}%
The BO energies $E_{N l m}^{\rm{BO}}(R)$ and the separation constants $\lambda_{N l m}(R)$ for some excited electronic states of $\rm{H}_2^{+}$ computed at large distances $R$. The numbers in parentheses denotes the powers of ten to be multiplied. All values are given in atomic units. The reference values are taken from Ref. \cite{Peek71}.
}
\begin{ruledtabular}
\renewcommand{\arraystretch}{1.0}
\begin{tabular}{lcl}
\textrm{state}& $R$ &
\multicolumn{1}{l}{\,$E_{N l m}^{\rm{BO}}$ (Present/Ref. \cite{Peek71})/  } \\
&&\,$\lambda_{N l m}$ (Present/Ref. \cite{Peek71})   \\
\colrule

$3d \sigma_g$ &50     & $-$1.262 673 906 996 215 539 487 577 052 488($-1$) \\  
            &       & $-$1.262 673 906 996($-1$)  \\
            &       & $-$7.579 801 889 086 814 118 202 538 621 711($+1$) \\            
            &       & $-$7.579 801 889 087($+1$) \\ 
            
            &90     & $-$1.253 802 060 869 790 541 149 187 331 225($-1$) \\
            &       & $-$1.253 802 060 870($-1$)   \\
            &       & $-$1.358 934 082 084 459 026 365 146 867 789($+2$) \\
            &       & $-$1.358 934 082 084($+2$)\\
          
$5g \pi_g$    &50     & $-$5.737 246 183 910 849 059 533 104 430 085($-2$) \\
           &       & $-$5.737 246 183 911($-2$) \\
           &       & $-$6.954 309 290 022 620 755 949 470 218 992($+1$) \\   
           &       & $-$6.954 309 290 023($+1$) \\
                     
           &100    &$-$5.600 302 812 696 510 402 376 796 685 615($-2$) \\
           &       &$-$5.600 302 812 697($-2$) \\ 
           &       &$-$1.368 102 002 800 595 095 361 298 816 688($+2$)  \\
           &       &$-$1.368 102 002 801($+2$) \\ 
        
           &150    &$-$5.575 405 175 595 630 988 817 932 376 624($-2$) \\        
           &       &$-$5.575 405 175 596($-2$) \\          
           &       &$-$2.036 608 739 009 452 820 868 095 110 508($+2$) \\           
           &       &$-$2.036 608 739 009($+2$) \\
\end{tabular}
\end{ruledtabular}
\end{table}


\begin{table}[ht]
\caption{\label{tab:HighE-Rmin}
Calculated position $R_0$ (a.u.) of the BO potential minimum of $\rm{H}_2^{+}$ for selected highly excited  ${}^{2}\Sigma$ electronic states in comparison with the data from Ref. \cite{Greene18}.
}
\begin{ruledtabular}
\renewcommand{\arraystretch}{0.8}
\begin{tabular}{ c c c }
state ($q,k$)  &  Ref. \cite{Greene18}  &  present \\
\colrule
($189$, $28$) & 46698  & 46360 \\    
($201$, $30$) &  52646 & 52495 \\    
($237$, $11$) &  62245 & 62226 \\
($233$, $17$) &  62709 & 62669 \\
($265$, $2$)  & 73152  &  73147 \\
($283$, $0$)  & 82444  & 82392 \\
\end{tabular}
\end{ruledtabular}
\end{table}


\vspace{1cm}

\begin{figure}[b]
\includegraphics [width=0.8\textwidth,angle=0]{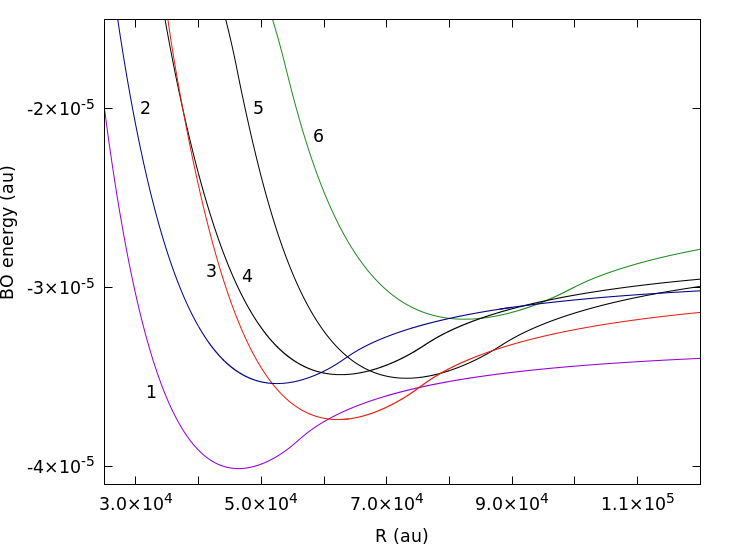}%
\caption{\label{fig:highE} PECs for selected highly excited states of $\rm{H}_2^{+}$ (see also Table \ref{tab:HighE-Rmin}): 1, $(q,k)=(189,28)$; 2, $(q,k)=(201,30)$; 3, $(q,k)=(237,11)$; 4, $(q,k)=(233,17)$; 5, $(q,k)=(265,2)$; 6, $(q,k)=(283,0)$.
}
\end{figure}

Solutions to the highly excited electronic states of $\rm{H}_2^{+}$ at very large internuclear separation has been the subject of the quite recent study \cite{Greene18}. The comparison of the calculated BO potential minimum $R_0$ for selected ${}^{2}\Sigma$ electronic states is given in Table \ref{tab:HighE-Rmin}. The overall agreement between the numbers is fairly well and is $\sim 0.5-0.03 \%$. 
For the state ($q=265$, $k=2$) our data agree to better than $\sim 0.007 \%$  with those reported in \cite{Greene18}. The computed potential energy curves for the considered electronic states are shown in Fig. \ref{fig:highE}.

\begin{figure}[b]
\includegraphics [width=0.8\textwidth,angle=0]{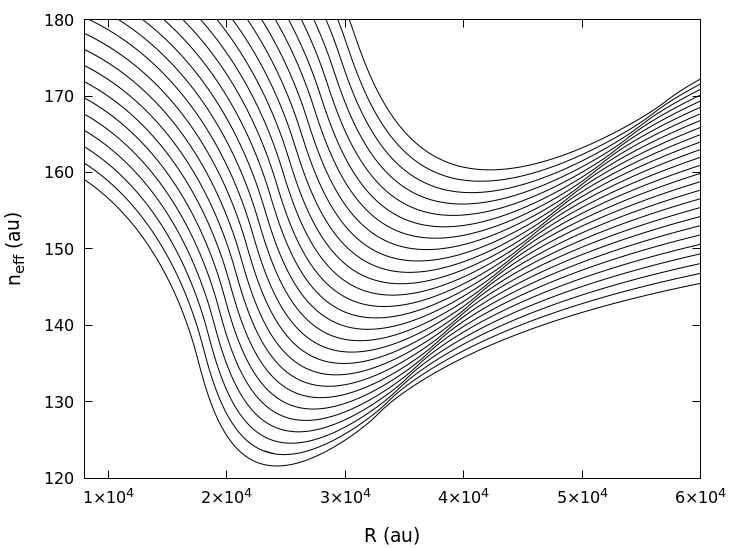}
\caption{\label{fig:ridge1} Ridge of avoided crossings shown for states $(k,q,m)=(0,q,0)$, $q=162,164,...,214$. 
}
\end{figure}

\begin{figure}[b]
\includegraphics [width=0.8\textwidth,angle=0]{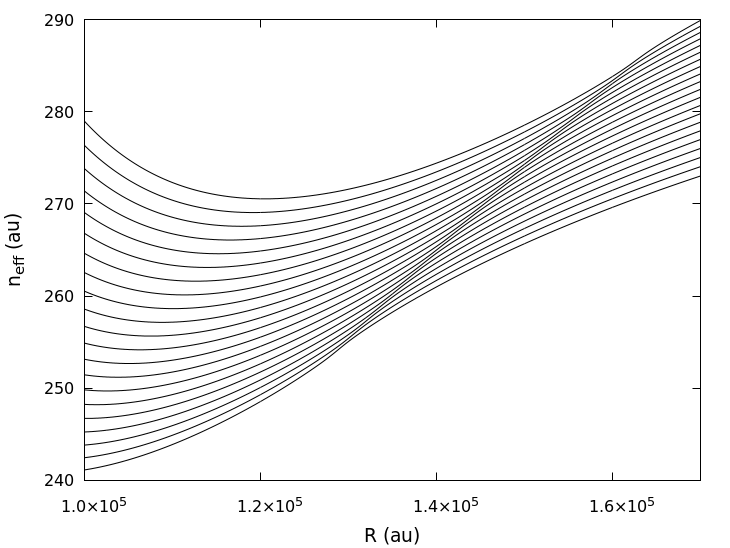}
\caption{\label{fig:ridge2} The same as in Fig. \ref{fig:ridge1} but for the states $q=322,324,...,362$.
}
\end{figure}

The value $n_{\rm eff}(R)=\sqrt{-2/E_{kqm}(R)}$ for the series of selected high-lying ${}^{2}\Sigma$ electronic states which undergo avoided crossings (i.e. with $k=0$ and $q=n,n+2,...$) is shown in Figures \ref{fig:ridge1} and \ref{fig:ridge2} for $q=$162$-$214 and $q=$322$-$362, respectively.
Our computed BO potentials demonstrate the typical behavior of the avoided crossings (see e.g., Fig.2a of the Ref. \cite{Greene18}) which also serves as a criterion for the correctness of the obtained results. We note, that the present computation algorithm is stable and can be routinely extended to even higher quantum numbers.

\clearpage

\subsection{Exact Demkov's solution}

It is known that there exist exact analytical solutions \cite{Demkov} of the $eZ_1Z_2$ problem for certain combinations of parameters $Z_1$, $Z_2$ ($Z_1 \neq Z_2$), and $R$. 
We have used as a benchmark test a one such solution for the $3p\sigma$ state for $Z_1 = 1$, $Z_2=5$, and $R=\sqrt{10}/3$ a.u. We have reproduced the analytic results for the electronic energy, $E=-2$ a.u., and the separation constant, $\lambda = -10/3$, with the precision of 32 decimal digits by setting $N_{\rm{RK}}=40$ and $N_{\rm{ NR}}=60$.

\subsection{Continuum spectrum of the two-Coulomb-center problem}\label{subsec:continuum}

The method presented here can also be applied to the computation of the continuum spectrum of the two-Coulomb-center problem, see Eqs. \eqref{def0-radial-eq} and \eqref{def0-angular-eq}. 
In fact, its application is much simpler than the one for the bound spectrum. For continuum spectrum the problem is reduced to finding the  eigenvalue $\lambda_{lm}(R)$ with the positive electronic energy $E$ treated as a free parameter. The expression \eqref{dEdR-res} is obviously not needed for this case and expression \eqref{dLdR} is simplified to
\begin{eqnarray}
\label{dLdR-cont}
\frac{d \lambda (R)}{dR}&& = -  { F_R^{(\eta)}}/
{ F_{\lambda}^{(\eta)} }\nonumber\\
&&=
-\frac{Z_{-} F_b^{(\eta)}  + (\kappa /2) F_c^{(\eta)}
} { F_{\lambda}^{(\eta)} },
\end{eqnarray}
were $c= \kappa R/2$ and $ \kappa =\sqrt{2E}$. For computation of the $F_R^{(\eta)}$, $F_{\lambda}^{(\eta)}$, 
$F_b^{(\eta)}$, and $F_c^{(\eta)}$, one can use the formulas \eqref{coef-ANG2} 
provided that the formal substitution, $p \to i c$, is made.

As an illustration, we present here the results of a calculation of the separation constant $\lambda_{00}$ for the electronic continuum of $\rm{HeH}^{2+}$ ($Z_1=1$ and $Z_2=2$). 
The Runge-Kutta propagation has been carried out with $N_{\rm RK}=200$ and $\Delta R = 0.0005$ a.u.
Clearly, the convergence of the numerical propagation is very good in the present (one-dimensional) case. Therefore, we did not employ a Newton-Raphson refinement. 
Our results for $\lambda_{00}$ are compared to the results of  \cite{Nakamura-CPC} and given in Table \ref{tab:nakamura}. For the technical details on the construction of the electronic wave function 
from the computed $\lambda_{lm}(R)$
we refer the readers to \cite{RankinThorson, Nakamura-CPC}.

\begin{table}[ht]
\caption{\label{tab:nakamura}
Calculated eigenvalue $\lambda_{00}$ for the $s \sigma$ continuum state of $\rm{HeH}^{2+}$ ($l=0$, $m=0$)
at $R=1.0$ a.u. and for different values of $\kappa$ compared with the data from Ref. \cite{Nakamura-CPC}. All values are given in atomic units.
}
\begin{ruledtabular}
\renewcommand{\arraystretch}{0.8}
\begin{tabular}{ l l }
$\kappa $  &  $\lambda_{00}$ (Present/Ref. \cite{Nakamura-CPC}) \\
\colrule
0.2 &  $-$0.164 017 311 895 064 441 002 855 443 \\
    &  $-$0.164 017 311 895 064 \\
1.0 &  $-$0.317 583 235 505 184 819 402 454 361 \\
    &  $-$0.317 583 235 505 185 \\    
5.0 &  $-$4.679 927 586 097 650 085 563 349 296 \\
    &  $-$4.679 927 586 097 65  \\
\end{tabular}
\end{ruledtabular}
\end{table}

\newpage

\subsection{SWE and GSWE with complex valued parameters}\label{subsec:complex}

Here we present a number of examples that illustrate the computation of the eigenvalues $\lambda_{lm}$ of equation \eqref{def0-angular-eq} if one allows for complex values of the parameters $b$ and $c$. 
Solving the spheroidal wave equations with complex valued parameters is generally a numerically difficult task \cite{CPC-spheroidal-2014,Falloon-03,Li-PRE-1998,Schmid-2023}. The advantage of the present approach is that, it can be directly applied for calculation of the eigenvalues both for SWE and GSWE with complex valued parameters.

In the present subsection we employed the expansion coefficients given by \eqref{coef-ANG1} with the formal substitution $p \to i c$. 
Furthermore, we use the same equation \eqref{dLdR-cont}
\[
\frac{d \lambda_{l m} (c,b|R)}{dR} = \frac{ (Z_1 - Z_2) F_b^{(\eta)}  - (\kappa /2) F_c^{(\eta)}
} { F_{\lambda}^{(\eta)} },
\]
where now the (generally complex) variable $R$ parametrizes the path $0 \to R_{f}$, 
and $Z_1$, $Z_2$, and $\kappa$ are the parameters to be defined
from solving the system of algebraic equations: i) $b=R_{f} (Z_2-Z_1)$ and ii) $c=\kappa R_{f} /2$.
The eigenvalue $\lambda_{l m}(c,b)$ can be obtained by integrating \eqref{dLdR-cont} along the path in the complex $R$-plane connecting two points $R=0$ and $R=R_f$. For instance, let $b=0.75+i$ and $c=1+0.5i$ then one may chose $(Z_2-Z_1)=0.5$, $\kappa=0.8-0.4i$, and $R_f=1.5+2i$. 
%
%
Computation of the eigenvalues $\lambda_{lm}$ for SWE (i.e. Eq. \eqref{def0-angular-eq} with $b=0$) is generally much simpler, it is sufficient to set $\kappa=2$, then $R_f$ is obviously equal to $c$.

Care must be given to the presence of the branch points $\bar{c}$ in the complex $c$-plain, when calculating the spectra of $\lambda_{lm}(c,b)$. If $c$ is large or located near branch point $\bar{c}$, then i) angular expansion \eqref{ANG1-expans} converges very slowly and ii) two different integration paths (in the complex $R$-plane) may lead to different numerical values for $\lambda(c)$ we are after, which corresponds to the different Riemann manifolds of $\lambda(c)$. Basically, the present algorithm allows to locate the positions of the branch points in $c$-plane. However, we will not address this problem here and 
refer the readers to publications \citep{Oguchi-RadioSci-70, Skorohodov-2006, Barrowes-2004} for more details. 

Our results of the calculations of eigenvalues for SWE  are presented in Table \ref{tab:Falloon1} (for real $c^2$) and Table \ref{tab:Falloon2} (for complex valued $c$) and compared with the results reported in \cite{CPC-spheroidal-2014, Falloon-03, Li-PRE-1998}. Overall, the present results are in better agreement with the data of \cite{Li-PRE-1998} than those reported in a more recent study \cite{CPC-spheroidal-2014}. The computed eigenvalues for SWE and GSWE for real $c$ are presented in Table \ref{tab:GSWE} and compared with the data from \citep{YanPengGong-PRE09} and \citep{Falloon-thesis}. The present results for the eigenvalues 
$\lambda_{lm}$ for GSWE with complex valued parameter $c=1+i$ and selected values of $b$ are given in Table \ref{tab:GSWE-complex-b}.


%
%

\begin{table}[ht]
\caption{\label{tab:Falloon1}
Comparison of eigenvalues $\lambda_{lm}(c,b=0)$ for real $c^2$ computed in the present work (PW) and in publications \cite{Falloon-03, Li-PRE-1998, CPC-spheroidal-2014}. 
}
\begin{ruledtabular}
\renewcommand{\arraystretch}{1.0}
\begin{tabular}{lllcll}
$m$& $l$&  $c$& Ref. &$-\lambda_{lm}(c)$ & $\lambda_{lm} ( {\rm i} c)$ \\ 
\colrule
0& 0& 1 &  PW & 0.68099 99448 53107 26021 60180 141  & 0.65139 76005 29730 91052 36276 172 \\
0& 0& 10 &  PW & $90.77169~ 57027~ 50054~ 84898~ 77312~ 426$ & $18.97205~ 60550~ 42243~ 81391~ 0919$  \\
 &  &    & \cite{Falloon-03} & $90.77169~ 57027~ 50054~ 84898~ 77312$   & $18.97205~ 60550~ 42243~ 81391~ 09191$  \\ 
 0& 0& 100& PW  & 9900.75189 88910 16747 44954 21523 33 & 198.99747 46340 82502  \\
 &  &    & \cite{Falloon-03}& 9900.75189 88910 16747 44954 21523  & 198.99747 46340 82548 13572 48103 \\ 
 1&  1& 1& PW\footnote{Result for $\lambda(c) + c^2$}  &  2.19554 83554 13003 95688 27437 346 & 1.79530 45872 81818 78854 10816 \\
 & & & \cite{Li-PRE-1998} & 2.19554 8355  & \\
 & & & \cite{CPC-spheroidal-2014} & 2.19561 23696 53500  & \\
1&  1&  4&  PW$^{\rm a}$ & 4.39959 30671 65506 10459 61890 349 
   & $-$2.90920 07591 46191 61848 25064 \\ 
&  & & \cite{Li-PRE-1998} & 4.39959 3067  & \\ 
&  &  & \cite{CPC-spheroidal-2014} & 4.39959 97606 64940  & \\
 1&  1&10& PW  & 89.71223 12326 08531 82924 20083 558 & 37.88064 98956 19453 22628 71048  \\   
 &   &   & \cite{Falloon-03} & 89.71223 12326 08531 82924 20083     & 37.88064 98956 19453 22628 71049 \\  
 1& 1& 100& PW & 9899.74682 23865 85061 62347 2435446 & 397.98984 67939 13  \\
 & & & \cite{Falloon-03} & 9899.74682 23865 85061 62347 24354 &  397.98984 67939 13121 45974 40124  \\
 4& 11& $i$ & PW$^{\rm a}$ &  131.56008 09194 06941 64691 87548   & \\
  & & & \cite{CPC-spheroidal-2014} & 131.56008 09183 03  & \\
\end{tabular}
\end{ruledtabular}
\end{table}

\begin{table}[ht]
\caption{\label{tab:Falloon2}
Comparison of eigenvalues  $\lambda_{00} (c,b=0)$ for complex parameter $c$ computed in the 
present work (PW) and in publications \cite{Falloon-03, CPC-spheroidal-2014}. Notation: $\alpha=1+i$.
}
\begin{ruledtabular}
\renewcommand{\arraystretch}{1.0}
\begin{tabular}{ r c l l}
 $c$ &  Ref.   & $ {\rm Re} [ \lambda_{00} (c) ]$ & $ {\rm Im} [ \lambda_{00} (c) ]$ \\ 
\colrule
 $\alpha$ & PW & 0.05947 27697 35031 26247 06156 230 & $-$1.33717 48778 05399 97103 72378 512  \\ 
 & \cite{Falloon-03} & 0.05947 27697 35031 26247 06156 & $-$1.33717 48778 05399 97103 72378  \\
$5\alpha$ & PW & 4.23035 06988 78380 87794 25891 & $-$44.97310 67423 02780 62895 65852  \\ 
$10\alpha$ & PW &  9.24076 62146 34603 35159 57442 763 & $-$189.98934 85956 57553 67515 08696 381   \\
 & \cite{Falloon-03} & 9.24076 62146 34603 35159 57442  & $-$189.98934 85956 57553 67515 08696 \\
 20$\alpha$ & PW\footnote{Result for $\lambda_{00}(c) + c^2$}  &  19.24532 81345 43127 94480   & 20.00499 41449 70279 83195  \\
 & \cite{CPC-spheroidal-2014} & 19.24532 81302 15245  &  20.00499 41471 04920 \\
 
\end{tabular}
\end{ruledtabular}
\end{table}

\begin{table}[ht]
\caption{\label{tab:GSWE}
Comparisons of eigenvalues $\lambda_{lm}(c) + c^2$ for selected values of $b$ computed in the present work (PW), Ref. \citep{YanPengGong-PRE09}, and Ref. \citep{Falloon-thesis} when available.
}
\begin{ruledtabular}
\renewcommand{\arraystretch}{1.0}
\begin{tabular}{ l l l l l}
$(m,l)$ &  $c$ & Ref. & $ b=0$   &   $b=2$ \\
\colrule

$(0,0)$&  1.0 & PW & 0.31900 00551 46892 73978 39819  & $-$0.15430 49702 80370 60175 78170  \\
 && \citep{YanPengGong-PRE09}  &  0.31900 00551 469  &   $-$0.15430 49702 80      \\   
 && \citep{Falloon-thesis}  &  0.31900 00551 46893  &  \\


& 25 & PW & 24.24209 35412 27324 02377 17920 &  24.24045 94381 87742 70543 68343 \\
&     & \citep{YanPengGong-PRE09} &  24.24209 35412 28   &   24.24045 94381 88     \\ 
%

& 100 & PW & 99.24810 11089 83252 55045 78476 &  99.24800 06013 19520 62785 79264 \\
&     & \citep{YanPengGong-PRE09} & 99.24810 11089 91           &       99.24800 06013 07  \\
&     & \citep{Falloon-thesis} & 99.24810 11089 832 &      \\
\colrule
$(4,8)$ & 1.0 & PW & 72.38941 89146 97508 96893 7413 &  72.39169 16942 19545 24437 8883 \\
 &   & \citep{YanPengGong-PRE09} &  72.38941 89146 98  &  72.39169 16942 19 \\
 &  25 & PW  & 233.57957 22097 14548 60892 011  & 233.57773 51174 30830 16340 670 \\
 & & \citep{YanPengGong-PRE09} &   233.57957 22097 1   &    233.57773 51174 3 \\
\end{tabular}
\end{ruledtabular}
\end{table}

\begin{table}[ht]
\caption{\label{tab:GSWE-complex-b}
The eigenvalues $\lambda_{lm}(c,b)$ for $c=1+i$ and selected values of the parameter $b$ computed in the present work.
}
\begin{ruledtabular}
\renewcommand{\arraystretch}{1.0}
\begin{tabular}{ c  l l l l l}
$m$,$l$  & $c$  & $b$  &~& Re [$\lambda_{lm}$] & Im [$\lambda_{lm} $]  \\ 
\colrule
(0,0) & $1+i$ & 0.01 &~& 0.05945 80409 82913 62742 87308 & $-$1.33716 78162 85459 23808 81316 \\

 & & 0.01$i$ &~& 0.05948 74985 33125 82592 62801 & $-$1.33718 19395 18751 81975 40535  \\

&  & 0.1+ 0.1$i$ &~& 0.05805 95356 20813 93544 20700 & $-$1.34011 67707 52950 85696 85414 \\

 & & 0.3+0.5$i$ &~&0.06081 32408 14327 63312 77640 
 &$-$1.39222 04483 35822 70100 69598  \\ 
 
\colrule

(0,1) & $1+i$ & 0 &~& 2.02747 01052 01366 13579 07339 & $-$0.79950 95706 03188 60956 42471 \\
 
 && 0.01 &~& 2.02747 79378 57379 30137 37533 & $-$0.79951 77153 27835 41634 07851  \\
 
 && 0.01$i$ &~& 2.02746 22724 93682 82343 60919 & $-$0.79950 14256 78483 84505 33634 \\ 
 
 && 0.1+0.1$i$ &~& 2.02910 00947 85808 27271 30658 & $-$0.79794 70293 39855 88573 06335 \\ 
 
 && 0.3+0.5$i$ &~& 2.04046 37161 43788 52379 97841 & $-$0.76341 32698 21135 99778 82071 \\ 

\end{tabular}
\end{ruledtabular}
\end{table}



\begin{figure}[b]
\includegraphics [width=0.8\textwidth,angle=0]{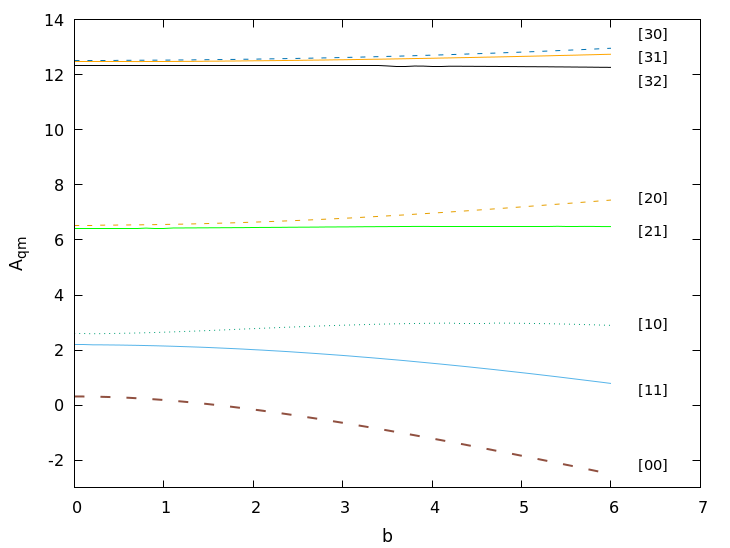}
\caption{\label{fig:gswe} The values $A_{qm} = \lambda_{qm} + c^2$ as a function of the parameter $b$ at $c=1$. The numbers in the square brackets $[qm]$ denotes the quantum numbers $q$ and $m$. 
}
\end{figure}

The results of the numerical calculations of the value $A_{qm}(c,b) = \lambda_{qm}(c,b) + c^2$ 
for GSWE at $c=1$ as the function of the parameter $b$ for selected quantum numbers $q$ and $m$ are shown in Fig. \ref{fig:gswe}.

\clearpage
\setlength{\parskip}{0pt}

\section{Concluding remarks}
\label{sec:conclusion}

We have presented an algorithm for solving the generalized spheroidal wave equations by employing the derivatives of the eigenvalues with respect to the distance $R$ between the foci of the spheroidal system of coordinates. The method of construction these derivatives in analytic form by means of continued fractions was given.
%
%
As an application of the developed approach, we have performed calculations of the eigenvalues (electronic energies and separation constants) for the two-Coulomb-center problem for several benchmark geometries of the system. We have demonstrated the application of the method both for the discrete and continuum spectrum of the $e Z_1 Z_2$ problem as well as to the solution for GSWEs with complex valued parameters. 
A representative set of results are presented and compared with the calculations of other authors. In the vast majority of cases, we obtained excellent agreement with the results of other authors, but in some cases a considerable disagreement is observed.  


To conclude, we confirm that proposed algorithm is robust and can be used for an exceptionally wide range of the parameters of the problem. The accuracy of the method is high and is actually limited by the accuracy of the employed computer arithmetic.





\appendix*
\section{}

The coefficients $\alpha^{(\xi)}_s$, $\beta^{(\xi)}_s$, and $\gamma^{(\xi)}_s$ for the expansion \eqref{Jaffe-expans} reads 
\[
\alpha^{(\xi)}_s = (s+1)(s+m+1),
\]
\[
\beta^{(\xi)}_s = -\lambda^{(\xi)} + 2 s \left( 2 p + s -\sigma \right) - 2p \sigma -(m+1)(m+\sigma),
\]
\begin{equation}\label{coef-rad-Jaffe}
\gamma^{(\xi)}_s = \left(s-\frac{a}{2 p}\right) \left( s-\frac{a}{2 p}+m \right).
\end{equation}

The coefficients $\alpha^{(\eta)}_s$, $\beta^{(\eta)}_s$, and $\gamma^{(\eta)}_s$ for the expansion \eqref{ANG1-expans} reads 
\[
\alpha^{(\eta)}_s = 
\frac{(s+ 2m +1)[b-2 p (s +m +1)] }{2 (s+m)+3},
\]
\[
\beta^{(\eta)}_s = \lambda^{(\eta)} + (s+m)(s+m+1),
\]
\begin{equation}\label{coef-ANG1}
\gamma^{(\eta)}_s = \frac{s (b+2 p (m+s))}{2 (m+s)-1}.
\end{equation}

For the basis $v_s^{+}$ (see Eq. \eqref{ANG2}) the coefficients $\alpha^{(\eta)}_s$, $\beta^{(\eta)}_s$, 
and $\gamma^{(\eta)}_s$ reads
\[
\alpha^{(\eta)}_s = 2(s+1)(s+m+1),
\]
\[
\beta^{(\eta)}_s =  \lambda^{(\eta)} + b +( 2s+m+1)(2 p +m) + s(s+1),
\]
\begin{equation}\label{coef-ANG2}
\gamma^{(\eta)}_s = b + 2p(s+m).
\end{equation}
For the basis $v_s^{-}$ the coefficients $\alpha^{(\eta)}_s$, $\beta^{(\eta)}_s$, 
and $\gamma^{(\eta)}_s$ are given by the same expressions \eqref{coef-ANG2}, albeit with the substitution $b \to -b$.

\begin{acknowledgments}

I thank Matt Eiles for many useful comments on the manuscript, 
Yurii Azhniuk for valuable comments on the Introduction part, and Eugen Remeta for many encouraging discussions. This study was partially supported by the U.S. Office of Naval Research Global (grant N62909-23-1-2088). The main part of the computations was undertaken with the assistance of resources from the National Computational Infrastructure (NCI), which is supported by the Australian Government in the framework of Ukraine-Australia Research Fund 2023.
The Visitors Program from the Max Planck Institute for the Physics of Complex Systems is greatly acknowledged. The financial support from the Krzysztof Skubiszewski Foundation on early stage of this research is greatly acknowledged.

\end{acknowledgments}





\input aps_ezz.bbl
\end{document}

%% file: aps_ezz.bbl
\providecommand{\noopsort}[1]{}\providecommand{\singleletter}[1]{#1}%